\documentstyle[12pt,amsfonts]{article}

\advance \textheight by \topskip
\makeatletter
\def\eqnarray{\stepcounter{equation}\let\@currentlabel=\theequation
\global\@eqnswtrue
\global\@eqcnt\z@\tabskip\@centering\let\\=\@eqncr
$$\halign to \displaywidth\bgroup\@eqnsel\hskip\@centering
  $\displaystyle\tabskip\z@{##}$&\global\@eqcnt\@ne
  \hfil$\displaystyle{\hbox{}##\hbox{}}$\hfil
  &\global\@eqcnt\tw@ $\displaystyle\tabskip\z@
  {##}$\hfil\tabskip\@centering&\llap{##}\tabskip\z@\cr}
\@addtoreset{equation}{section}
  \def\theequation{\thesection.\arabic{equation}}
\makeatother
\def\mbar#1{\kern 0.1em\overline{\kern -0.1em #1 \kern -0.1em} \kern 0.1em}
\def\Ps#1{\raisebox{.2ex}{$\displaystyle
  \mathop{\Psi }^{\scriptscriptstyle #1}$}{}}
\def\Ph#1{\raisebox{.2ex}{$\displaystyle
  \mathop{\Phi }^{\scriptscriptstyle #1}$}{}}

\begin{document}

\begin{titlepage}
\hbox to \hsize{\hfil hep-th/9404119}
\hbox to \hsize{\hfil INR 0848/94}
\hbox to \hsize{\hfil March, 1994}
\vfill
\large \bf
\begin{center}
GENERALIZED SCHR\"ODINGER REPRESENTATION \\
IN BRST--QUANTIZATION
\end{center}
\vskip 1cm
\normalsize
\begin{center}
{\bf Kh.S. Nirov}\footnote{E--mail: nirov@inucres.msk.su}\\
{\small \it Institute for Nuclear Research of the Russian Academy of
Sciences} \\
{\small \it 60th October Anniversary prospect 7a, 117312 Moscow, Russia}\\
and \\
{\bf A.V. Razumov}\footnote{E--mail: razumov@mx.ihep.su} \\
{\small \it Institute for High Energy Physics, 142284 Protvino,
Moscow Region, Russia}
\end{center}
\vskip 2.cm
\begin{abstract}
\noindent
An analysis of the state space in the BRST--quantization in the
Schr\"odinger representation is performed on the basis of the results
obtained earlier in the framework of the Fock space representation.
It is shown that to get satisfactory results it is necessary to have from
the very beginning a meaningful definition of the total state space.
\end{abstract}
\vfill
\end{titlepage}

\section{Introduction}

Now the BRST--quantization method is the most popular method for the
covariant quantization of gauge--invariant systems. This method is based
on the concept of the BRST--symmetry, being a special type of symmetry
generated by a nilpotent operator $\widehat \Omega$ \cite{BRS74}. The
procedure of the BRST--quantization in its simplest form looks as follows
\cite{KuU82}. Starting with a given gauge--invariant system, characterized
by a gauge--invariant Lagrangian, we extend the configuration space by
adding ghost and antighost variables, and construct a BRST--invariant
effective Lagrangian. This effective Lagrangian, unlike the original one,
is nondegenerate.  The usual methods of quantization applied to the
effective Lagrangian leads us to the total state space of the
BRST--quantization, where the BRST--symmetry generator $\widehat \Omega$
acts.  The physical subspace is specified then by the condition
\begin{equation}
\widehat \Omega \Psi = 0. \label{1.1}
\end{equation}
Since we have
\begin{equation}
\widehat \Omega^2 = 0, \label{1.2}
\end{equation}
then the vectors of the form
\begin{equation}
\Psi = \widehat \Omega \Phi \label{1.3}
\end{equation}
evidently belong to the physical subspace. Moreover, such vectors are
orthogonal to any physical vector. Factorizing the physical subspace by
the subspace, formed by the vectors of form (\ref{1.3}), we come to the
physical state space of the system.

The BRST--charge $\widehat \Omega$ is by definition a hermitian operator.
It is quite clear that a hermitian nilpotent operator in a space with a
positive definite scalar product is trivial. Hence, the total state space
must have an indefinite scalar product, and the question on positivity of
the scalar product in the physical state space arises. In fact, the
complete answer to this question is yet unknown. For a special class of
systems with a quadratic BRST--charge a strict proof of the positivity of
the scalar product in the physical state space was given in
Ref.~\cite{RaR90}. The proof was based on the representation of the total
state space as a ${\Bbb Z}_2$--graded Krein space. The relevant operators
were there creation and annihilation operators. There were a few attempts
to analyze the structure of the total state space starting with the
operators of generalized coordinates and momenta in the Schr\"odinger
representation, but these attempts have not led, from our point of view,
to satisfactory results. In the present paper we translate the results of
Ref.~\cite{RaR90} to the language of the Schr\"odinger representation and
discuss the drawbacks of the previous considerations.

The main difference between the consideration given here and the attempts
made earlier is indefiniteness of the scalar product in the sector of the
bosonic unphysical operators. This obstacle leads to the necessity of
using the so--called generalized Schr\"odinger representation for the
operators of generalized coordinates and momenta \cite{AFIO81}. The next
very important point, allowing to get satisfactory results, is the usage
of an auxiliary positive definite scalar product in the total state space.
This scalar product defines the norm which allows to get rid of the
superfluous physical states arising in a naive treatment of the problem.

\section{Formulation of the problem}

There are two different approaches to constructing a BRST--invariant
effective theory starting from a given gauge--invariant system. In the
first approach, discussed in the introduction, the initial object is a
gauge--invariant Lagrangian. Following the second approach, one starts
with the Hamiltonian description of the initial system. The constraints,
arising in this description, determine the form of the corresponding
BRST--charge (for a review see Refs.~\cite{Hen85,BaF86}). In fact, both
approaches are essentially equivalent (see Ref.~\cite{NiR93} and
references therein).  In the present consideration we are interested only
in the form of the BRST--charge and follow the Hamiltonian approach.

Let us consider the hamiltonian system with the phase space described by
generalized coordinates $q_i$, $Q_\alpha$, $i = 1, \ldots, n$, $\alpha =
1,\ldots, m$, and generalized momenta $p_i$, $P_\alpha$, having the usual
nonzero Poisson brackets:
\begin{equation}
\{p_i, q_j\} = -\delta_{ij}, \qquad  \{P_\alpha, Q_\beta\} =
-\delta_{\alpha \beta}. \label{2.1}
\end{equation}
Suppose that there are $n$ first class constraints \cite{Dir58} of the form
\begin{equation}
p_i = 0. \label{2.2}
\end{equation}
To construct the BRST--charge \cite{Hen85,BaF86}, associated with the
system under consideration, we enlarge the phase space of the system adding
to the initial (even) coordinates odd coordinates $\theta_i$, $\pi_i$, $i
= 1, \ldots, n$, with the nonzero Poisson brackets
\begin{equation}
\{\pi_i, \theta_j\} = - \delta_{ij}. \label{2.4}
\end{equation}
According to the general scheme
\cite{Hen85,BaF86}, the BRST--charge in our case
has the form
\begin{equation}
\Omega = \theta_i p_i. \label{2.5}
\end{equation}

We shall treat the commuting and anticommuting variables as even and odd
elements of a Grassmann algebra ${\Bbb G}$ with countably infinite number
of generators over the field of complex numbers ${\Bbb C}$
\cite{Rog80,DeW84,PRR88}.  Suppose that the Grassmann algebra ${\Bbb G}$ is
supplied with an involution, so that we can define the concept of a real
variable. It is natural to consider the even variables $q_i$, $Q_\alpha$,
$p_i$ and $P_\alpha$ as real variables. We suppose also that the odd
generalized coordinates $\theta_i$ are real, while the odd generalized
momenta $\pi_i$ are imaginary \cite{PRR88}. Denoting the involution in
${\Bbb G}$ by an asterisk, we can write
\begin{eqnarray}
&q^*_i = q_i, \qquad p^*_i = p_i, \qquad
Q^*_\alpha = Q_\alpha, \qquad P^*_\alpha = P_\alpha, & \label{2.6} \\
&\theta^*_i = \theta_i, \qquad \pi^*_i = -\pi_i.& \label{2.7}
\end{eqnarray}

In quantum theory we have the set of the operators $\widehat q_i$,
$\widehat Q_\alpha$, $\widehat \theta_i$, corresponding to the generalized
coordinates, and the operators $\widehat p_i$, $\widehat P_\alpha$,
$\widehat \pi_i$, corresponding to the generalized momenta. These
operators irreducibly act in a Hilbert space $\cal H$ with an indefinite
scalar product $(\ ,\ )$.  According to Eqs.~(\ref{2.6}), (\ref{2.7}) we
suppose that
\begin{eqnarray}
&\widehat q^\dagger _i = \widehat q_i, \qquad \widehat p^\dagger _i =
\widehat p_i, \qquad \widehat Q^\dagger _\alpha = \widehat Q_\alpha,
\qquad \widehat P^\dagger _\alpha = \widehat P_\alpha,& \label{2.8} \\
&\widehat \theta^\dagger _i = \widehat \theta_i, \qquad \widehat
\pi^\dagger _i = - \widehat \pi_i,& \label{2.9}
\end{eqnarray}
where $\dagger$ means the hermitian conjugation with respect to the scalar
product $(\ ,\ )$.  The operators, we have introduced, satisfy the
following commutation relations
\begin{equation}
[\widehat p_i, \widehat q_j] = - i \delta_{ij}, \qquad [\widehat P_\alpha,
\widehat Q_\beta] = -i \delta_{\alpha \beta}, \qquad [\widehat \pi_i,
\widehat \theta_j] = -i \delta_{ij}, \label{2.10}
\end{equation}
where only the nontrivial relations are written. Note that the symbol $[\
,\ ]$ means here the generalized commutator \cite{RaR90}. According to the
classical expression, we suppose that the BRST--charge in quantum theory
has the form
\begin{equation}
\widehat \Omega = \widehat \theta_i \widehat p_i. \label{2.11}
\end{equation}
It is clear that the operator $\widehat \Omega$ is hermitian and nilpotent.

It is convenient to consider a concrete representation of the operators we
are dealing with. The most popular here is the Schr\"odinger
representation. In this representation the state space $\cal H$ of the
system under consideration is formed by the functions of commuting
variables $q_a$, $Q_\alpha$ and anticommuting variables $\theta_a$, having
the form
\begin{equation}
\Psi(q, Q, \theta) = \sum_{k=0}^n \frac{1}{k!} \Ps{(k)}_{i_1 \ldots i_k}
(q, Q) \, \theta_{i_1} \ldots \theta_{i_k}, \label{2.12}
\end{equation}
where the functions $\Ps{(k)}_{i_1 \ldots i_k}(q, Q)$ take values in
${\Bbb G}$. Actually we restrict ourself to the functions $\Ps{(k)}_{i_1
\ldots i_k}(q, Q)$ being continuations of ordinary functions of $n+m$ real
variables \cite{DeW84}.

We define the integral of a function $\Psi(q, Q, \theta)$ over commuting
and anticommuting variables in accordance with Ref.~\cite{DeW84}, with the
only difference that the integration over the odd variables $\theta_i$ is
normalized by
\begin{equation}
\int \Psi(q, Q, \theta) d^n \theta = \Ps{(n)}_{1 \ldots n}(q, Q).
\label{2.13}
\end{equation}
The scalar product in $\cal H$ is defined as
\begin{equation}
(\Psi, \Phi) = (-i)^{n(n-1)/2} \int \Psi^*(q, Q, \theta) \Phi(q, Q,
\theta) d^n q d^m Q d^n \theta. \label{2.14}
\end{equation}
The operators, corresponding to the generalized coordinates, are realized
in the Schr\"odinger representation as multiplication operators:
\begin{equation}
\widehat q_i \Psi  = q_i \Psi,  \qquad
\widehat Q_\alpha \Psi = Q_\alpha \Psi, \qquad
\widehat \theta_i \Psi = \theta_i \Psi,
\label{2.16}
\end{equation}
while the operators, corresponding to the generalized momenta, are
proportional to differentiation operators:
\begin{equation}
\widehat p_i \Psi = -i \frac{\partial \Psi }{\partial q_i}, \qquad
\widehat P_\alpha \Psi = -i \frac{\partial \Psi }{\partial Q_\alpha},
\qquad \widehat \pi_i \Psi = -i \frac{\partial \Psi }{\partial \theta_i}.
\label{2.18}
\end{equation}
It can be shown that such definition of the operators and the scalar
product leads us to relations (\ref{2.8})--(\ref{2.10}).

For the action of the BRST--charge on an arbitrary state vector $\Psi(q,
Q, \theta)$ we get the expression
\begin{equation}
(\widehat \Omega \Psi)(q, Q, \theta) = -i \sum_{k=0}^{n-1}
\frac{1}{k!} \left(\widehat p_{[i_1} \Ps{(k)}_{i_2 \ldots
i_{k+1}]}\right)(q, Q) \,
\theta_{i_1} \ldots \theta_{i_{k+1}}, \label{2.19}
\end{equation}
where the square brackets means the antisymmetrization.  From the above
expression it follows that the BRST--invariance condition (\ref{1.1}) is
equivalent in our case to the set of relations
\begin{equation}
\widehat p_{[i_1} \Ps{(k)}_{i_2 \ldots i_{k+1}]} = 0,
\qquad k = 0, \ldots, n-1. \label{2.20}
\end{equation}
In particular, for $k = 0,1$ we have
\begin{eqnarray}
&\widehat p_i \Ps{(0)} = 0,& \label{2.21} \\
&\widehat p_i \Ps{(1)}_j - \widehat p_j \Ps{(1)}_i = 0.& \label{2.22}
\end{eqnarray}
{}From Eq.~(\ref{2.22}) it follows that there exists a function $\Ph{(0)}(q,
Q)$ such that
\begin{equation}
\Ps{(1)}_i = \widehat p_i \Ph{(0)}. \label{2.23}
\end{equation}
In fact, it can be shown that there exists a set of functions
$\Ph{(k)}_{i_1 \ldots i_k}(q, Q)$, such that
\begin{equation}
\Ps{(k+1)}_{i_1 \ldots i_{k+1}} = (k+1) \widehat p_{[i_1} \Ph{(k)}_{i_2
\ldots i_{k+1}]}, \qquad k = 0, \ldots n-1. \label{2.24}
\end{equation}
Thus, any BRST--invariant state vector can be represented in the form
\begin{equation}
\Psi = \Ps{(0)} + \widehat \Omega \Phi, \label{2.25}
\end{equation}
where the function $\Ps{(0)}(q, Q)$ satisfies relations (\ref{2.21}). To
prove the validity of this statement, it is convenient to establish an
analogy of the action of the operator $\widehat \Omega$ with the action of
the exterior derivative operator. To this end, let us treat the variables
$Q_\alpha$ as parameters, and associate with a state vector of form
(\ref{2.12}) the differential form
\begin{equation}
\widetilde \Psi = \sum_{k=0}^n \frac{1}{k!} \Ps{(k)}_{i_1\ldots i_k}(q, Q)
\, dq_{i_1} \wedge \cdots \wedge dq_{i_k}. \label{2.26}
\end{equation}
It is not difficult to get convinced that
\begin{equation}
\widetilde{\widehat \Omega \Psi} = -i d \widetilde \Psi. \label{2.27}
\end{equation}
Using now the Poincar\'e lemma \cite{Sch82}, we easily come to
representation (\ref{2.25}).

Factorizing out the state vectors of form (\ref{1.3}), we conclude that
the physical state space is formed by functions which do not depend on
$q_i$ and $\theta_i$. This result is, at a first sight, very attractive,
because it establishes a direct correspondence between the physical state
space in the BRST--quantization and the state space, arising in the Dirac
quantization of the system \cite{Dir58}. From the other side, we see that
the scalar product for the physical state vectors is undefined. We can get
for it either zero or infinite value depending on the order of the
integration over the commuting and anticommuting variables we shall
choose. This discouraging result forces us to consider the above
reasonings more carefully.

The first observation, we can make, is that we have not actually given a
strict definition of the total state space. What functions of commuting
and anticommuting variables do really belong to it? Recall that in a usual
situation, which we encounter in quantum mechanics, we consider as the
vectors, describing the states of a system in the Schr\"odinger
representation, the square integrable functions. For such functions the
scalar product takes finite values. In our case the situation is more
complicated. If we define the state space as the set of vectors having
finite scalar square, we shall discover that not any pair of such vectors
has finite scalar product.

In any case a strict definition of the state space may destroy the
consideration given above, because it may happen that the vector $\Phi$,
entering representation (\ref{2.25}), does not belong to the state space,
and the physical state space is different from that we have gotten in the
above consideration.

Recall also that the functions, which do not depend on $q_i$ and
$\theta_i$, correspond to the state vectors which trivially satisfy the
BRST--invariance condition, and the scalar product for such states is
undefined. Hence, any reasonable definition of the state space should
allow to get rid of such vectors.

Note here that in Ref.~\cite{RaR90} there was given a strict consideration
of the problem we are discussing here. The main difference from the present
paper was the usage in Ref.~\cite{RaR90} of creation and annihilation
operators in the Fock space representation instead of operators of
generalized coordinates and momenta in the Schr\"odinger representation
which we use here. It is interesting to translate the results of
Ref.~\cite{RaR90} to the language of the Schr\"odinger representation and
compare them with the results obtained here.

\section{State space of model}

Let us begin with the discussion of the state space of the model
considered in Ref.~\cite{RaR90}. First recall that the classical
formulation of the BRST--quantization involves both commuting and
anticommuting variables, furthermore it is supposed that an involution for
these variables is introduced. Accordingly, in quantum theory the
corresponding operators act in a ${\Bbb Z}_2$--graded Hilbert space, and
the involution corresponds to the hermitian conjugation in this space.

For the model considered in Ref.~\cite{RaR90} there are the operators
$A_\alpha$ and $A^\dagger _\alpha$ corresponding to physical particles,
and the operators $a_i$, $\mbar a_i$, $c_i$, $\mbar c_i$ and $a^\dagger
_i$, $\mbar a^\dagger _i$, $c^\dagger _i$, $\mbar c^\dagger _i$
corresponding to unphysical particles. It is supposed that the operators
of physical and unphysical particles act in a ${\Bbb Z}_2$--graded Hilbert
space ${\cal H}$. The scalar product in ${\cal H}$ is denoted by $\langle\
,\ \rangle$, and the hermitian conjugate operator of an operator $O$ is
denoted by $O^\dagger $. It is assumed that
\begin{equation}
|A_\alpha| = \mbar 0, \qquad |a_i| = |\mbar a_i| = \mbar 0, \qquad |c_i| =
|\mbar c_i| = \mbar 1. \label{3.1}
\end{equation}
Here and below we follow the notations of Ref.~\cite{RaR90}. The
nontrivial commutation relations for the operators of physical and
unphysical particles have the form
\begin{eqnarray}
&[A_\alpha, A_\beta^\dagger ] = \delta_{\alpha \beta},& \label{3.2} \\
&[a_i, \mbar a_j^\dagger ] = \delta_{ij}, \qquad [\mbar a_i, a^\dagger _j]
= \delta_{ij},& \label{3.3} \\
&[c_i, \mbar c^\dagger _j] = \delta_{ij}, \qquad [\mbar c_i, c^\dagger _j]
= \delta_{ij}.& \label{3.4}
\end{eqnarray}
In these relations, $[\ ,\ ]$ denotes the generalized commutator
\cite{RaR90}. The form of the commutation relations allows us to call the
operators with a dagger and the operators without it, creation and
annihilation operators, respectively.

Suppose that in ${\cal H}$ there exists a unique vacuum vector $\Psi_0$,
such that
\begin{eqnarray}
&A_\alpha\Psi_0 = 0,& \label{3.5} \\
&a_i \Psi_0 = \mbar a_i \Psi_0 = 0,& \label{3.6} \\
&c_i \Psi_0 = \mbar c_i \Psi_0 = 0,& \label{3.7} \\
&\langle\Psi_0, \Psi_0 \rangle = 1.& \label{3.8}
\end{eqnarray}
Due to a non--canonical form of the commutation relations (\ref{3.3}) and
(\ref{3.4}) the state space of the system is an indefinite metric space.
It is worth to note here that the indefiniteness of the metric is
connected not only with the odd ghost operators $c_i$ and $\mbar c_i$, but
also with the even operators $a_i$ and $\mbar a_i$ corresponding to the
gauge degrees of freedom. To demonstrate the indefiniteness of the metrics
it is convenient to introduce a new set of even unphysical operators
defined by
\begin{eqnarray}
&a_{0i} = \frac{1}{\sqrt 2}(\mbar a_i - a_i), \qquad a_{0i}^\dagger  =
\frac{1}{\sqrt 2} (\mbar a_i^\dagger  - a_i^\dagger ),& \label{3.9} \\
&a_{1i} = \frac{1}{\sqrt 2}(\mbar a_i + a_i), \qquad a_{1i}^\dagger  =
\frac{1}{\sqrt 2} (\mbar a_i^\dagger  + a_i^\dagger ).& \label{3.10}
\end{eqnarray}
As it follows from Eq.~(\ref{3.6}) the operators $a_{0i}$ and $a_{1i}$
annihilate the vacuum vector:
\begin{equation}
a_{0i} \Psi_0 = a_{1i} \Psi_0 = 0. \label{3.11}
\end{equation}
The nontrivial commutation relations for the new operators have the form
\begin{equation}
[a_{0i}, a^\dagger _{0j}] = - \delta_{ij}, \qquad [a_{1i}, a^\dagger
_{1j}] = \delta_{ij}. \label{3.12}
\end{equation}
Consider the vector $a^\dagger _{0i}\Psi_0$; from relations (\ref{3.11})
and (\ref{3.12}) we see that the scalar square of this vector is equal to
$-1$. Thus, we actually deal with the state space with indefinite metric.

As it was noted in \cite{RaR90}, the consideration of the state space in
the BRST--quantization becomes rigorous if we treat it as a ${\Bbb
Z}_2$--graded Krein space \cite{Bog74,AzI86,RaR90}. Let us describe the
corresponding construction. Consider the Fock space ${\cal H}$ where the
operators $A_\alpha$, $a_i$, $\mbar a_i$, $c_i$, $\mbar c_i$ act as
ordinary annihilation operators. Here $A_\alpha$, $a_i$, $\mbar a_i$ are
boson annihilation operators, while $c_i$, $\mbar c_i$ are fermion
annihilation operators. As it was noted above the space ${\cal H}$ should
be a ${\Bbb Z}_2$--graded linear space, with $A_\alpha$, $a_i$, $\mbar
a_i$ being even operators, and $c_i$, $\mbar c_i$ being odd operators. We
introduce in ${\cal H}$ the corresponding structure of a ${\Bbb
Z}_2$--graded linear space as follows.

Denote the positive definite scalar product in ${\cal H}$ by $(\ ,\ )$.
The hermitian conjugation with respect to this scalar product will be
denoted by a star. Hence, $A_\alpha^\star$, $a_i^\star$, $\mbar
a_i^\star$, $c_i^\star$, $\mbar c_i^\star$ are the creation operators,
corresponding to the annihilation operators $A_\alpha$, $a_i$, $\mbar
a_i$, $c_i$, $\mbar c_i$. The annihilation and creation operators are
supposed to satisfy the ordinary commutation or anticommutation relations.
The vectors of ${\cal H}$, generated by the action of the creation
operators on the vacuum vector $\Psi_0$ form a basis in ${\cal H}$. Let us
assume that a vector generated by an even (odd) number of fermion
creation operators and any number of boson creation operators is even
(odd). In this the vacuum vector $\Psi_0$ is considered to be even. An
arbitrary vector of ${\cal H}$ is said to be even (odd) if it can be
represented as a linear combination of even (odd) basis vectors. It is
clear that we have actually introduced in ${\cal H}$ the required
structure of ${\Bbb Z}_2$--graded Hilbert space \cite{RaR90}. Taking into
account the fact that in a ${\Bbb Z}_2$--graded Hilbert space the parity
of the hermitian conjugate operator coincides with the parity of the
original one, we can write the commutation relation of the creation and
annihilation operators with the help of the generalized commutator
operation, as
\begin{eqnarray}
&[A_\alpha, A_\beta^\star] = \delta_{\alpha \beta},& \label{3.13} \\
&[a_i, a_j^\star] = \delta_{ij}, \qquad [\mbar a_i, \mbar a_j^\star] =
\delta_{ij},& \label{3.14} \\
&[c_i, c_j^\star] = \delta_{ij}, \qquad [\mbar c_i, \mbar c_j^\star] =
\delta_{ij}.& \label{3.15}
\end{eqnarray}

Introduce now in ${\cal H}$ the structure of a ${\Bbb Z}_2$--graded Krein
space.  To this end, define the operator $J$ with the help of the
relations
\begin{eqnarray}
&J A_\alpha J^{-1} = A_\alpha,& \label{3.16} \\
&J a_i J^{-1} = \mbar a_i, \qquad J \mbar a_i J^{-1} = a_i,&
\label{3.17} \\
&J c_i J^{-1} = \mbar c_i, \qquad J \mbar c_i J^{-1} = c_i,&
\label{3.18} \\
& J \Psi_0 = \Psi_0.& \label{3.19}
\end{eqnarray}
It can be easily shown that the operator $J$ is a hermitian operator,
satisfying the relation $J^2 = I$. Note also that $J$ is an even operator.
Thus, $J$ defines in ${\cal H}$ the structure of a ${\Bbb Z}_2$--graded
Krein space \cite{RaR90}. The corresponding indefinite scalar product
$\langle\ ,\ \rangle$ is related to the positive definite scalar product
$(\ ,\ )$ by the equality
\begin{equation}
\langle \phi, \psi \rangle \equiv (\phi, J\psi). \label{3.20}
\end{equation}
Note also that the hermitian conjugation with respect to the indefinite
scalar product is related to the hermitian conjugation with respect to the
positive definite scalar product by the relation
\begin{equation}
O^\dagger  = JO^\star J \label{3.21}
\end{equation}
for any operator $O$. From this relation and from Eqs~(\ref{3.16})--(3.18)
it follows that
\begin{eqnarray}
&A_\alpha^\dagger  = A_\alpha^\star,& \label{3.22} \\
&a_i^\dagger  = \mbar a_i^\star, \qquad \mbar a_i^\dagger  = a_i^\star,&
\label{3.23} \\
&c_i^\dagger  = \mbar c_i^\star, \qquad \mbar c_i^\dagger  = c_i^\star.
\label{3.24}
\end{eqnarray}
Commutation relations (\ref{3.2})--(\ref{3.4}) is now a consequence of
commutation relations (\ref{3.13})--(\ref{3.15}). The introduction of the
Krein space allows us to use in the consideration of the state space,
arising in the BRST--quantization, ordinary methods of investigation of
the Hilbert spaces with positive definite scalar product. In particular,
the concept of the Krein space can be used to prove the positive
definiteness of the scalar product in the physical state space for the
case of the quadratic BRST--charge \cite{RaR90}.

Without any loss of generality, we can suppose that the greek and latin
indices take just one value. We assume that this is the case and denote
the corresponding operators by the same letters without indices.

\section{Generalized Schr\"odinger representation}

{}From the consideration of the previous section it follows that we can
consider the state space ${\cal H}$ as the tensor product of three spaces
${\cal H}_a$, ${\cal H}_A$ and ${\cal H}_c$:
\begin{equation}
{\cal H} = {\cal H}_a \otimes {\cal H}_A \otimes {\cal H}_c.
\end{equation}
The space ${\cal H}_a$ is the representation space for the even unphysical
creation and annihilation operators, the space ${\cal H}_c$ is the
representation space for the odd unphysical creation and annihilation
operators, while ${\cal H}_A$ is the corresponding space for the physical
operators. Here the operator $J$ and the vacuum vector $\Psi_0$ factorize
as
\begin{eqnarray}
&J = J_a \otimes J_A \otimes J_c,& \\
&\Psi_0 = \Psi_{0a} \otimes \Psi_{0A} \otimes \Psi_{0c}.&
\end{eqnarray}

\subsection{Even unphysical operators}

First consider the space ${\cal H}_a$. Introduce the operators $\widehat
v_r$ and $\widehat u_r$, $r = 0,1$ defined by
\begin{equation}
\widehat v_r = \frac{1}{\sqrt 2} (a_r + a_r^\star), \qquad \widehat u_r =
\frac{1}{i\sqrt 2} (a_r - a_r^\star). \label{4.1}
\end{equation}
The operators $\widehat v_r$, $\widehat u_r$ are hermitian with respect to
the positive definite scalar product $(\ ,\ )$:
\begin{equation}
\widehat v_r^\star = \widehat v_r, \qquad \widehat u_r^\star = \widehat
u_r \label{4.2}
\end{equation}
and satisfy the commutation relations
\begin{eqnarray}
&[\widehat v_r, \widehat v_s] = 0, \qquad [\widehat u_r, \widehat u_s] =
0,& \label{4.3} \\
&[\widehat v_r, \widehat u_s] = i \delta_{rs}.& \label{4.4}
\end{eqnarray}
Thus, we can consider ${\cal H}_a$ as the space formed by square integrable
function of two real variables $v_0$ and $v_1$ with the scalar product
\begin{equation}
(\Psi, \Phi) = \int \Psi^*(v) \Phi(v) d^2 v. \label{4.5}
\end{equation}
The Schr\"odinger representation for the operators $\widehat v_r$ and
$\widehat u_r$ is
\begin{equation}
\widehat v_r \Psi = v_r \psi, \qquad \widehat u_r \Psi = -i \frac{\partial
\Psi}{\partial v_r}. \label{4.6}
\end{equation}
Hence, for the annihilation and creation operators we have
\begin{equation}
\widehat a_r \Psi = \frac{1}{\sqrt 2} \left(v_r + \frac{\partial}{\partial
v_r}\right) \Psi, \qquad \widehat a^\star_r \Psi = \frac{1}{\sqrt 2}
\left(v_r - \frac{\partial}{\partial v_r}\right) \Psi. \label{4.7}
\end{equation}
The vacuum vector $\Psi_{0a}$ in the space ${\cal H}_a$ has the form
\begin{equation}
\Psi_{0a}(v) = \frac{1}{\sqrt \pi} e^{- \frac{1}{2}(v_0^2 + v_1^2)}.
\label{4.8}
\end{equation}

Let us consider the action of the operator $J_a$ on ${\cal H}_a$. From the
definition of the operators $a_r$ we get
\begin{equation}
J_a a_0 J_a = - a_0, \qquad J_a a_1 J_a = a_1. \label{4.9}
\end{equation}
{}From these relations it follows that
\begin{eqnarray}
&J_a \widehat v_0 J_a = - \widehat v_0, \qquad J_a \widehat u_0 J_a = -
\widehat u_0,& \label{4.10} \\
&J_a \widehat v_1 J_a = \widehat v_1, \qquad J_a \widehat u_1 J_a =
\widehat u_1. \label{4.11}
\end{eqnarray}
These equalities, together with the condition of the invariance of the
vacuum vector $\Psi_{0a}$ give
\begin{equation}
(J \Psi)(v_0, v_1) = \Psi (-v_0, v_1). \label{4.12}
\end{equation}
We can write this result shortly as
\begin{equation}
(J \Psi)(v) = \Psi(\sigma(v)), \label{4.13}
\end{equation}
where $\sigma(v_0, v_1) = (-v_0, v_1)$.  For the indefinite scalar product
we get the expression
\begin{equation}
\langle \Psi, \Phi \rangle = \int \Psi^*(\sigma(v)) \Phi(v) d^2 v.
\label{4.14}
\end{equation}

The operators $\widehat v_0$ and $\widehat u_0$ are not hermitian with
respect to the indefinite scalar product $\langle \ ,\ \rangle$; actually
we have
\begin{equation}
\widehat v_0^\dagger = - \widehat v_0, \qquad \widehat u_0^\dagger = -
\widehat u_0. \label{4.15}
\end{equation}
Since the indefinite scalar product is more fundamental for our problem,
it is desirable to introduce new operators, which are hermitian with
respect to it \cite{Mar93a}. To this end, let us consider the operators
$\widehat q_r$ and $\widehat p_r$, defined by
\begin{eqnarray}
&\widehat q_0 = - i \widehat v_0, \qquad \widehat p_0 = i \widehat u_0,&
\label{4.16} \\
&\widehat q_1 = \widehat v_1, \qquad \widehat p_1 = \widehat u_1.
\label{4.17}
\end{eqnarray}
For these operators we have
\begin{equation}
\widehat q_r^\dagger = \widehat q_r, \qquad \widehat p_r^\dagger =
\widehat p_r.
\end{equation}
The commutation relations are again of the canonical form
\begin{eqnarray}
&[\widehat q_r, \widehat q_s] = 0, \qquad [\widehat p_r, \widehat p_s] =
0,& \label{4.18} \\
&[\widehat q_r, \widehat p_s] = i \delta_{rs}.& \label{4.19}
\end{eqnarray}

Introducing the variables
\begin{equation}
q_0 = -i v_0, \qquad q_1 = v_1, \label{4.20}
\end{equation}
we can consider ${\cal H}_a$ as the space formed by functions of these
variables. For the operators $\widehat q_r$
and $\widehat p_r$ we have
\begin{equation}
\widehat q_r \Psi = q_r \Psi, \qquad \widehat p_r \Psi = -i\frac{\partial
\Psi}{\partial q_r}, \label{4.21}
\end{equation}
while the expressions for the scalar products takes the form
\begin{eqnarray}
&(\Psi, \Phi) = i \int \Psi^*(q) \Phi(q) d^2 q,& \label{4.22} \\
&\langle \Psi, \Phi \rangle = i \int \Psi^*(q^*) \Phi(q) d^2 q,&
\label{4.23}
\end{eqnarray}
where $d^2 q = dq_0 dq_1 = -i dv_0 dv_1$.

\subsection{Odd unphysical operators}

Introduce now the operators of the odd generalized coordinates $\widehat
\theta_r$ and generalized momenta $\widehat \pi_r$:
\begin{eqnarray}
&\widehat \theta_0 = \frac{1}{\sqrt 2}(c^\dagger + c), \qquad \widehat
\theta_1 = \frac{1}{i\sqrt 2}(c - c^\dagger),& \label{4.24} \\
&\widehat \pi_0 = \frac{1}{i\sqrt 2}(\mbar c + \mbar c^\dagger), \qquad
\widehat \pi_1 = \frac{1}{\sqrt 2} (-\mbar c + \mbar c^\dagger).
\label{4.25}
\end{eqnarray}
The operators $\widehat \theta_r$ and $\widehat \pi_r$ satisfy the
canonical commutation relations
\begin{eqnarray}
&[\widehat \theta_r, \widehat \theta_s] = 0, \qquad [\widehat \pi_r,
\widehat \pi_s] = 0,& \label{4.26} \\
&[\widehat \pi_r, \widehat \theta_s] = - i \delta_{rs}.& \label{4.27}
\end{eqnarray}
Hence we can take as the space ${\cal H}_c$ the space of functions of two
real anticommuting variables $\theta_r$, $r = 0,1$, and define the
operators $\widehat \theta_r$ and $\widehat \pi_r$ as
\begin{equation}
\widehat \theta_r \Psi = \theta_r \Psi, \qquad \widehat \pi_r \Psi =
- i \frac{\partial \Psi}{\partial \theta_r}. \label{4.28}
\end{equation}
The indefinite scalar product in this case have the form
\begin{equation}
\langle \Psi, \Phi \rangle = -i \int \Psi^*(\theta) \Phi(\theta) d^2
\theta. \label{4.29}
\end{equation}
It is not difficult to get convinced that the operators $\widehat
\theta_r$ are hermitian, while the operators $\widehat \pi_r$ are
antihermitian with respect to the scalar product $\langle\ ,\ \rangle$.

The annihilation operators $c$ and $\mbar c$ act on a state vector as
\begin{equation}
c \Psi = \frac{1}{\sqrt 2}(\theta_0 + i\theta_1) \Psi, \qquad \mbar c \Psi
= \frac{1}{\sqrt 2}\left( \frac{\partial}{\partial \theta_0} + i
\frac{\partial}{\partial \theta_1}\right) \Psi. \label{4.30}
\end{equation}
Hence the vacuum vector $\Psi_{0c}$ has the form
\begin{equation}
\Psi_{0c}(\theta) = \frac{1}{\sqrt 2}(\theta_0 + i \theta_1) \label{4.31}
\end{equation}

{}From the definition of the operators $\widehat \theta_r$ and $\widehat
\pi_r$ it follows that
\begin{equation}
J_c \widehat \theta_r J_c = i \widehat \pi_r, \qquad J_c \widehat \pi_r
J_c = -i \widehat \theta_r. \label{4.32}
\end{equation}
Taking into account the invariance of the vacuum vector under the action
of the operator $J_c$, we can get the following expression for the action
of $J_c$ on an arbitrary state vector:
\begin{equation}
J_c \Psi(\theta) = -i \int e^{(\theta_0 \eta_0 + \theta_1 \eta_1)}
\Psi(\eta) d^2 \eta \label{4.33}
\end{equation}
{}From this relation we conclude that the positive definite scalar product
in ${\cal H}_c$ has the form
\begin{equation}
(\Psi, \Phi) = - \int \Psi^*(\theta) e^{(\theta_0 \eta_0 + \theta_1
\eta_1)} \Phi(\eta) d^2 \theta d^2 \eta. \label{4.34}
\end{equation}

\subsection{Physical operators and total state space}

There are no problems with the construction of the Schr\"odinger
representation for the operators of physical particles. Introducing the
operators of the generalized coordinate $\widehat Q$ and generalized
momentum $\widehat P$:
\begin{equation}
\widehat Q = \frac{1}{\sqrt 2}(A + A^\star), \qquad \widehat P =
\frac{1}{i\sqrt 2}(A - A^\star), \label{4.35}
\end{equation}
we consider ${\cal H}_A$ as the space of square integrable function of a
real variable $Q$, and represent $\widehat Q$ and $\widehat P$ as
\begin{equation}
\widehat Q \Psi = Q \Psi, \qquad \widehat P \Psi = -i \frac{\partial
\Psi}{\partial Q}. \label{4.36}
\end{equation}
The operator $J_A$ acts in ${\cal H}_A$ as the unit operator, and the
scalar products $(\ ,\ )$ and $\langle\ ,\ \rangle$ coincide:
\begin{equation}
(\Psi, \Phi) = \langle \Psi, \Phi \rangle = \int \Psi^*(Q) \Phi(Q) dQ.
\label{4.37}
\end{equation}

Summarizing the above consideration, we can say that the total state space
is the space formed by functions of the variables $q_r$, $Q$ and
$\theta_r$. All these variables are real, except the variable $q_0$, which
is imaginary. The positive definite scalar product in the total state
space has the form
\begin{equation}
(\Psi, \Phi) = -i \int \Psi^*(q, Q, \theta) e^{(\theta_0 \eta_0 + \theta_1
\eta_1)} \Phi(q, Q, \eta) d^2 q d Q d^2 \theta d^2 \eta, \label{4.38}
\end{equation}
while for the indefinite scalar product we have the expression
\begin{equation}
\langle \Psi, \Phi \rangle = \int \Psi^*(q^*, Q, \theta) \Phi(q, Q,
\theta) d^2 q dQ d^2 \theta. \label{4.39}
\end{equation}
The operator $J$, connecting the indefinite and positive definite scalar
products, act on a state vector $\Psi$ as
\begin{equation}
(J \Psi)(q, Q, \theta) = -i \int e^{(\theta_0 \eta_0 + \theta_1 \eta_1)}
\Psi(q^*, Q, \eta) d^2 \eta. \label{4.40}
\end{equation}

\section{BRST--charge and physical state space}

It can be shown that a general quadratic BRST--charge can be written as
\cite{KuO79,Sla89,Ryb91}
\begin{equation}
\widehat \Omega = \sqrt 2(\mbar a^\dagger c + c^\dagger \mbar a).
\label{5.1}
\end{equation}
Proceeding to the operators of the generalized coordinates and momenta we
get for $\widehat \Omega$ the following expression
\begin{equation}
\widehat \Omega = (\widehat p_0 + \widehat q_1) \widehat \theta_0 +
(\widehat p_1 + \widehat q_0) \widehat \theta_1. \label{5.2}
\end{equation}
In Ref.~\cite{RaR90} it was shown that in the case under consideration any
BRST--invariant state vector can be represented in the form
\begin{equation}
\Psi = \Psi' + \widehat \Omega \Phi, \label{5.3}
\end{equation}
where the vector $\Psi'$ satisfy the relations
\begin{equation}
a \Psi' = \mbar a \Psi' = 0, \qquad c \Psi' = \mbar c \Psi' = 0. \label{5.4}
\end{equation}
In other words, the vector $\Psi'$ does not contain unphysical particles.
Note that the vector $\Psi'$ is defined by the vector $\Psi$ uniquely.
Moreover, different physical vectors, satisfying relations (\ref{5.4}),
correspond to different physical states.  Using the Schr\"odinger
representation, we get for $\Psi'$ the expression
\begin{equation}
\Psi'(q, Q, \theta) = \frac{1}{\sqrt{2 \pi}} (\theta_0 + i \theta_1) e^{
\frac{1}{2}(q_0^2 - q_1^2)} \psi(Q). \label{5.5}
\end{equation}
Thus, the physical state space can be parameterized by the square integrable
functions $\psi(Q)$.

The BRST--charge, given by Eq.~(\ref{5.2}), is different from the
BRST--charge, we considered in section 2. Note that after the
transformation
\begin{equation}
\widehat p_0 + \widehat q_1 \to \widehat p_0, \qquad \widehat p_1 +
\widehat q_0 \to \widehat p_1 \label{5.6}
\end{equation}
we come to the BRST--charge of form (\ref{2.11}). This transformation does
not change commutators, and after it we get the following Schr\"odinger
representation for the operators $\widehat q^r$ and $\widehat p_r$:
\begin{eqnarray}
&\widehat q_0 \Psi = q_0 \Psi, \qquad \widehat q_1 \Psi = q_1 \Psi,&
\label{5.7} \\
&\widehat p_0 \Psi = \left( -i\frac{\partial}{\partial q_0} + q_1\right)
\Psi, \qquad \widehat p_1 \Psi = \left( -i\frac{\partial}{\partial q_1} +
q_0 \right) \Psi.& \label{5.8}
\end{eqnarray}
To get the usual representation we should multiply the state vector by the
factor $\exp(-iq_0q_1)$. After this we obtain for the vector $\Psi'$ the
following expression
\begin{equation}
\Psi'(q, Q, \theta) = \frac{1}{\sqrt{2 \pi}} (\theta_0 + i \theta_1) e^{
\frac{1}{2}(q_0 - iq_1)^2} \psi(Q). \label{5.9}
\end{equation}
Note here that the multiplication of the state vectors by the factor
$\exp(-iq_0q_1)$ does not change the expression for the indefinite scalar
product $\langle \ ,\ \rangle$, while for the positive definite scalar
product we get
\begin{equation}
(\Psi, \Phi) = -i \int \Psi^*(q, Q, \theta) e^{(\theta_0 \eta_0 + \theta_1
\eta_1)} e^{2iq_0q_1} \Phi(q, Q, \eta) d^2 q d Q d^2 \theta d^2 \eta.
\label{5.10}
\end{equation}

Let us now compare the results we have obtained with the discussion given
in section 2. Formally we have the same Schr\"odinger representation as in
section 2. The difference is in the fact that the variable $q_0$ is now
imaginary and the scalar product have form (\ref{4.39}). Besides, we also
have a positive definite scalar product defined on the state space.
Actually the total state space ${\cal H}$ in our approach is defined as
the space of the functions $\Psi$ of the variables $q_r$, $Q$ and
$\theta_r$, such that $(\Psi, \Psi) < \infty$.  The arguments based on the
Poincar\'e lemma are also applicable in our case, but taking into account
the results of Ref.~\cite{RaR90}, we conclude that we cannot now transform
a general physical state vector to the state vector having no dependence
from odd variables. The physical state vectors having no dependence from
odd variables, should be excluded from the consideration, because the
positive definite scalar product is not defined for them.

\section{Conclusion}

In the present paper we have compared the consideration of the state space
in the BRST--quantization made in Ref.~\cite{RaR90} with the help of
creation and annihilation operators in the Fock representation, with one
based on the operators of generalized coordinates and momenta in the
Schr\"odinger representation.  It appeared  that the naive treatment of
the problem leads to unsatisfactory results due to the absence of a
meaningful definition of the total phase space.

Stress also that in accordance with our consideration some variables
describing the initial gauge--invariant system should be quantized with
indefinite metric, this conclusion is in accordance with a general
consideration of the properties of the BRST--quantization performed by R.
Marnelius with collaborators (see Ref.~\cite{Mar93b} and references
therein).

Unfortunately, we do not see now a way to generalize the results obtained
in the present paper to the case of constraints forming a general
nonabelian algebra. Its seems very likely that such a generalization
should be based on the consideration of the structure of the corresponding
gauge group manifold, but it is not clear for us how to introduce an
indefinite metrics in the space of functions on the gauge group. This
question is quite nontrivial, and a short remark made in this respect in
Ref.~\cite{Mar93b} is, from our point of view, not enough.
\vskip 0.5cm
One of the authors (Kh.N.) is indebted to Profs. V.A. Rubakov and
F.V. Tkachov for support and discussions.  His research was supported
in part by the Weingart Foundation through a cooperative agreement with
the Department of Physics at UCLA. This work was also supported by the
program of short--term research grants of the Soros International Science
foundation.

\end{document}